\documentclass[reprint,prb,twocolumn,showpacs,superscriptaddress,aps,longbibliography]{revtex4-2}
 
\usepackage{natbib}
\usepackage[colorlinks=true,citecolor=blue]{hyperref} % hyperreferences
\usepackage{graphicx}
\usepackage{bm}
\usepackage{amsmath}
\usepackage{amssymb}
\usepackage{physics}
\UseRawInputEncoding
 
\DeclareGraphicsExtensions{.png,.jpg,.eps}
\usepackage{xcolor}
\usepackage{orcidlink}

\begin{document}
 
\preprint{APS/123-QED}

\title{\textbf{Self-consistent tensor network method for correlated super-moir\'e matter beyond one billion sites} 
}% 

\author{Yitao Sun\,
\thanks{yitao.sun@aalto.fi}
\orcidlink{0009-0002-9479-7147}}
\author{Marcel Niedermeier\,\orcidlink{0000-0002-0988-6464}}
 \author{Tiago V. C. Antao\,\orcidlink{0000-0003-3622-2513}}
\author{Adolfo O. Fumega\,\orcidlink{0000-0002-3385-6409}}
\author{Jose L. Lado\,\orcidlink{0000-0002-9916-1589}}

\affiliation{
 Department of Applied Physics, Aalto University, 02150 Espoo, Finland
}

\date{\today}% It is always \today, today,

\begin{abstract}
Moir\'e and super-moir\'e materials provide exceptional platforms to engineer exotic correlated
quantum matter. 
The vast number of sites required to model moir\'e
systems in real space remains a formidable challenge due to the immense computational resources required. 
Super-moir\'e materials push this requirement to the limit, where millions or even billions of sites need to be considered, a requirement beyond the capabilities of conventional methods for interacting systems.
Here, we establish a methodology that allows solving correlated states in systems reaching a billion sites, that exploits
tensor network representations of real space Hamiltonians and self-consistent
real space mean-field equations.
Our method combines a tensor network kernel polynomial method with
quantics tensor cross interpolation algorithm,
enabling us to solve ultra-large models, including those
whose single-particle Hamiltonian is too large to be stored explicitly. 
We demonstrate our methodology with super-moir\'e systems
featuring spatially modulated hoppings, many-body interactions and domain walls, showing 
that it allows access to self-consistent symmetry-broken states and spectral functions
of real space models reaching a billion sites.
Our methodology provides a strategy to solve exceptionally large interacting problems,
providing a widely applicable strategy to compute correlated super-moir\'e quantum matter.
\end{abstract}

\maketitle

\section{Introduction}
Twisted moir\'e materials have enabled creating of a wide variety of correlated states of matter, including
unconventional superconducting phases~\cite{Park2021,Lu2019,Zhang2022,Yankowitz2019,Uri2023,Klein2023}, 
topological states~\cite{Zeng2023,Cai2023,Serlin2020,Zeng2023,Lu2024} and correlated insulators~\cite{Cao2018b,Zhao2023,Kim2023,Vao2021,Zhao2023}. These emergent states stem from the
moir\'e pattern between twisted or mismatched 2D materials~\cite{Andrei2021},
ideas that have been extended to a variety of artificial
platform including optical metamaterials~\cite{Fu2020,Du2023,Tang2023,Mao2021,PhysRevLett.130.083801,PhysRevLett.127.163902,Luan2023} and cold atom systems~\cite{PhysRevA.100.053604,PhysRevLett.125.030504,Yu_2024,Meng_2023}. 
The moir\'e pattern gives rise
to an emergent periodicity and associated unit cells ranging from thousands to 
hundreds of thousands of sites,
reaching the limits of theoretical atomistic methods to study correlated phases~\cite{PhysRevLett.119.107201,PhysRevLett.127.026401,PhysRevLett.115.106601,Joo2022,PhysRevB.107.125423,Carr2020}.
\begin{figure}[ht!]
    \centering
    \includegraphics[width=0.9\linewidth]{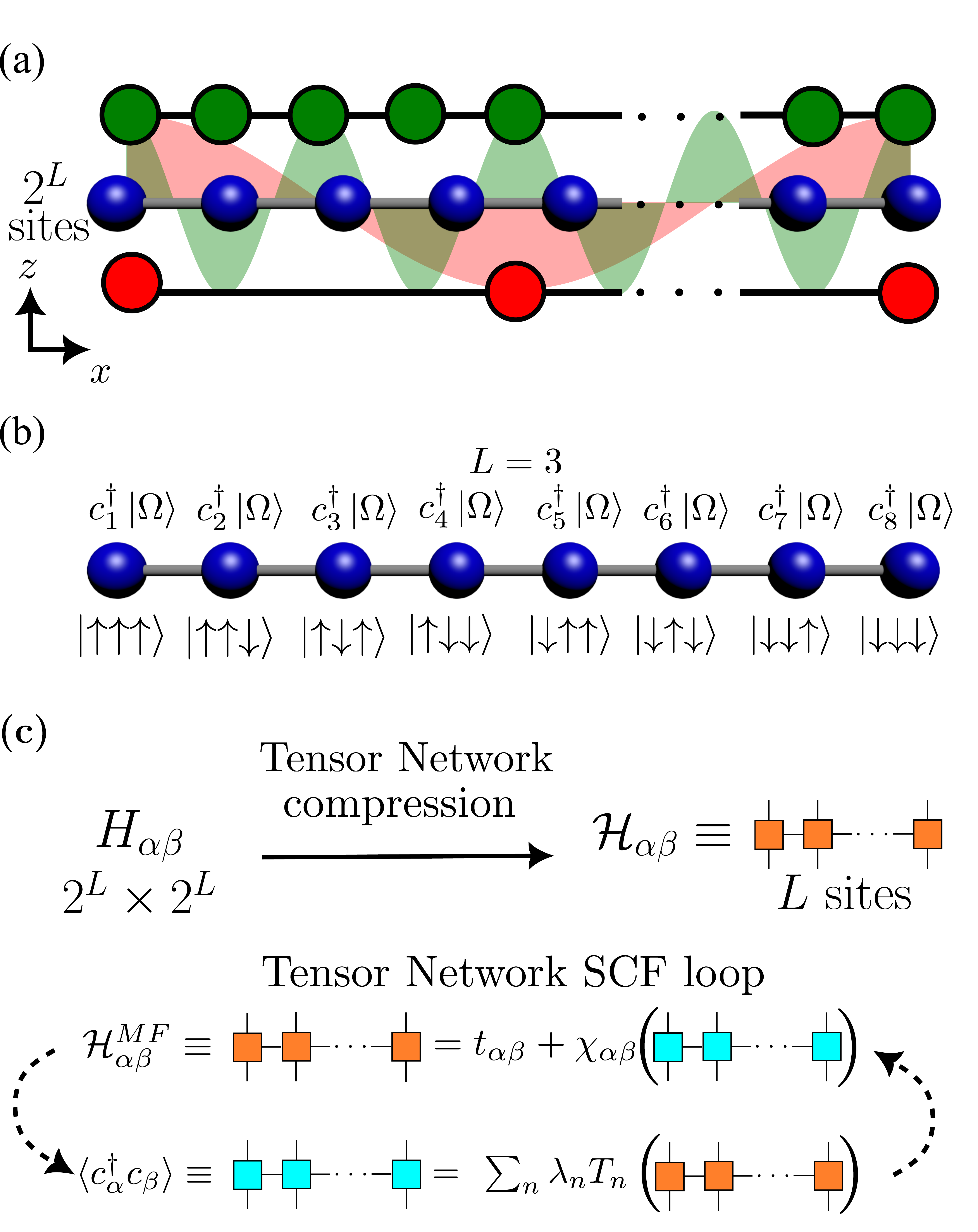}
   
    \caption{\textbf{Schematic of the tensor-network self-consistent algorithm for super-moir\'e matter.}
    Panel (a) shows a schematic of a system featuring two moir\'e patterns due to the mismatch
    of a top and bottom substrate. Panel (b) shows an example of representing electron states in fermionic and pseudo-spin bases. Panel (c) shows the tensor-network compression and mean-field algorithm, where
    the fermionic problem is mapped to a pseudo-spin problem.}
    %\vspace{-10pt}
    \label{fig:1}
\end{figure}
When two or more moir\'e patterns coexist, a super-moir\'e pattern emerges,
leading to an even wider variety of exotic phases~\cite{Li2022,Kapfer2023}. 
However, super-moir\'e materials require treating systems from millions to hundreds of millions
of sites, which is a challenge for conventional computational methods.

In many-body physics, tensor network algorithms have provided an exceptional method to
treat systems with ultra-large many-body Hilbert spaces~\cite{PhysRevLett.69.2863,RevModPhys.77.259,Schollwck2011,Ors2019}. 
The essence of tensor networks is to compress a Hilbert space of dimension
$2^L$ with a linearly scaling variational product of tensors.
This strategy has provided the most accurate solutions to paradigmatic quantum many-body problems~\cite{Yan2011,PhysRevB.92.041105,Xu2024,Jiang2019},
and has been recently extended to tackle problems in machine learning~\cite{PhysRevResearch.4.043007,PhysRevX.8.031012,NIPS2016_5314b967}
%, quantum chemistry~\cite{Nakatani2013,Chan2016,Szalay2015}
, quantum computing~\cite{PhysRevResearch.6.033325,PhysRevX.10.041038,Huang2021}
%, quantum error correction~\cite{Pastawski2015,PhysRevLett.127.040507,PhysRevA.105.052446} 
and ultimately general problems that require dealing with ultra-large structures of data and integrals~\cite{fernández2024learningtensornetworkstensor,PhysRevLett.132.056501,PhysRevX.13.021015, PhysRevB.107.245135,PhysRevX.12.041018,10.21468/SciPostPhys.18.1.007,PhysRevB.110.035124,Gourianov2022,Peddinti2024,Gourianov2025}.
Therefore, the capability of tensor networks to deal with ultra-large systems may provide a way forward to tackle problems of interacting
super-moir\'e materials~\cite{Fumega2024}.

Here, we establish a tensor network methodology to solve correlated states in super-moir\'e
systems reaching a billion sites.
Our method uses an atomistic self-consistent mean-field approach where both single-particle and interacting
mean-field terms are expressed purely as tensor networks,
using a combination of tensor network Chebyshev expansions and quantics tensor cross interpolation.
Our strategy leverages a mapping between the mean-field Hamiltonian of an exponentially
large super-moir\'e system to a linearly scaling pseudo-spin many-body problem.
This methodology allows for the treatment of super-moir\'e systems whose mean-field Hamiltonian is too large to be explicitly stored, enabling us to deal with system sizes beyond the capabilities of techniques with sparse matrix representations.
We demonstrate our method in several super-moir\'e structures with over one billion atomic sites, showing
that it allows us to solve self-consistently interacting problems with spatially varying hoppings and interactions.
This work thus provides a methodology capable of dealing with the exceptional system sizes required for
correlated phases in super-moir\'e quantum matter. 
The paper is organized as follows.
In Section~\ref{SCF}, we briefly introduce the mean-field algorithm employed in this work.
Section~\ref{construction} illustrates the construction of the tensor-network representation of the tight-binding Hamiltonian using the hopping Hamiltonian of a one-dimensional model as an example.
The numerical results for mean-field calculations over various one- and two-dimensional super-moir\'e systems are presented in Section~\ref{results}, together with a discussion of the advantages and possible extensions of our methodology.
Finally, Section~\ref{conclusion} summarizes the main findings of this study.

\section{Self-consistent tensor network algorithm} 
\label{SCF}
We consider a generic interacting real-space Hamiltonian
of a super-moir\'e system with $N=2^L$ sites as shown in Fig.~\ref{fig:1}(a), that has both single-particle and density-density
many-body interaction
taking the form
\begin{equation}
H = 
H_0 + H_V =\sum_{\alpha\beta} t_{\alpha\beta} c^\dagger_{\alpha} c_{\beta} + 
\sum_{\alpha\beta} V_{\alpha\beta} c^\dagger_{\alpha} c_{\alpha} c^\dagger_{\beta} c_{\beta},
\end{equation}
where $t_{\alpha\beta}$ are single-particle hopping amplitudes in the system and $V_{\alpha\beta}$ 
account for the many-body density-density interactions. 
The previous Hamiltonian can be solved at the mean-field level
by performing the mean-field decoupling $H_V \approx H^{MF}_V $
where
$
H^{MF}_V 
= \sum_{\alpha\beta}
V_{\alpha\beta} \langle c^\dagger_{\alpha} c_{\alpha} \rangle c^\dagger_{\beta} c_{\beta}
+ ... = \sum_{\alpha\beta} \chi_{\alpha \beta} c^\dagger_{\alpha} c_{\beta},
$
$...$ denoting the remaining Wick contractions.
This decoupling gives rise to a mean-field Hamiltonian of the form
\begin{equation}
H^{MF} = \sum_{\alpha \beta} (t_{\alpha \beta} + 
\chi_{\alpha \beta} )c^\dagger_{\alpha} c_{\beta}
= \sum_{\alpha \beta} H^{MF}_{\alpha \beta} c^\dagger_{\alpha} c_{\beta},
\label{eq:mf}
\end{equation}
with $H^{MF}_{\alpha\beta} \equiv H^{MF}_{\alpha\beta}  \left [V_{\alpha\beta},\langle c^\dagger_{\alpha} c_{\beta} \rangle \right ] $ 
parameterizing the self-consistent mean-field Hamiltonian
including single-particle and many-body corrections.
For large systems beyond millions of sites, the mean-field Hamiltonian $H^{MF}$
becomes too large to be even stored. We can now reinterpret the $2^L$-site
mean-field Hamiltonian as a many-body operator of an auxiliary many-body pseudo-spin chain of $L$ sites,
for which the index $\alpha,\beta$ corresponds to an element of the pseudo-spin many-body basis 
as $ \alpha \equiv (s_1,s_2,..,s_L)$
and
$ \beta \equiv (s'_1,s'_2,..,s'_L)$. In this form, the mean-field Hamiltonian
can be written using a tensor network representation as a matrix product operator (MPO) in the pseudo-spin basis as
\begin{equation}
\mathcal{H}^{MF}_{\alpha\beta} \equiv \Gamma_{s_1,s_1'}^{(1)} \Gamma_{s_2,s_2'}^{(2)} \Gamma_{s_3,s_3'}^{(3)}...\Gamma^{(L)}_{s_L,s_L'},
\end{equation}
where $\Gamma_{s_1,s_1'}^{(n)}$ are tensors of dimension $m$, the bond dimension of the MPO.
The structured nature of a super-moir\'e Hamiltonian enables the representation of $H^{MF}_{\alpha\beta}$ with
a manageable bond dimension~\cite{fernández2024learningtensornetworkstensor, PhysRevLett.132.056501}. The main problem to solve the interacting system is then to find the tensor network representation $\mathcal{H}^{MF}$ of $H^{MF}$ in Eq.~\ref{eq:mf}, which can be done by solving a Chebyshev tensor-network self-consistent
equation~\cite{itensor,Quantics.jl,TensorCrossInterpolation.jl,QuanticsTCI.jl,Besard2019,Besard20192}. The essential objects to compute are
correlators
$
\langle c^\dagger_{\alpha} c_{\beta} \rangle 
= 
\int_{-\infty}^{\epsilon_F}
\langle \alpha| \delta(\omega - \mathcal{H}^{MF}) |\beta \rangle
d\omega ,
$
with the matrix product state (MPS) $|\alpha \rangle = c^\dagger_\alpha |\Omega \rangle$ as shown in Fig.~\ref{fig:1}(b), $|\Omega\rangle$ the vacuum state,
$\epsilon_F$ the Fermi energy and $\delta(\omega - \mathcal{H}^{MF})$ Dirac delta function operator.
We can now compute all the expectation values as
\begin{equation}
\langle c^\dagger_{\alpha} c_{\beta} \rangle =
\langle \alpha| \Xi(\mathcal{H}^{MF}) |\beta \rangle,
\label{eq:cij}
\end{equation}
where $\Xi (\mathcal{H}^{MF})$ is the tensor-network representation of the density matrix
computed with a tensor-network Chebyshev kernel polynomial method (KPM)~\cite{PhysRevB.83.195115,PhysRevResearch.1.033009,PhysRevLett.130.100401,PhysRevResearch.6.043182,2025arXiv250605230A}. By rescaling the bandwidth of $\mathcal{H}^{MF}$ to the interval $(-1,1)$, the density matrix
takes the form
\begin{equation}
\Xi (\mathcal{H}^{MF}) = \sum_n \lambda_n T_n(\mathcal{H}^{MF}),
\label{eq:kpm}
\end{equation}
where $T_{n}(x)$ are the Chebyshev polynomials with
recursion relation $T_{n}(x) = 2xT_{n-1}(x) - T_{n-2}(x)$
with $T_0=1$ and $T_1(x) =x$~\cite{RevModPhys.78.275}.
The expansion coefficients are
$\lambda_n = \int_{-1}^{\epsilon_F} 
\frac{2T_{n}(\omega)}{\pi \sqrt{1-\omega^{2}}}d\omega
$ for $n\ge 1$ and
$\lambda_0 = \int_{-1}^{\epsilon_F} 
\frac{T_{0}(\omega)}{\pi \sqrt{1-\omega^{2}}}d\omega
$.
These Chebyshev recursion relations allow us to find a tensor network
representation of Eq.~\ref{eq:kpm} by performing
sums and contractions between MPOs.
While Eq.~\ref{eq:kpm} is not of practical use 
with large sparse matrix representations, 
it can be directly applied with MPOs~\footnote{The expansion in Eq.~\ref{eq:kpm} 
can be formally applied with conventional sparse matrices $H^{MF}$, however, the full matrix with elements $\langle c^\dagger_{\alpha} c_{\beta} \rangle$ will be dense even when $H^{MF}$ is sparse}.

\begin{figure}[t!]
    \centering
    \includegraphics[width=1\linewidth]{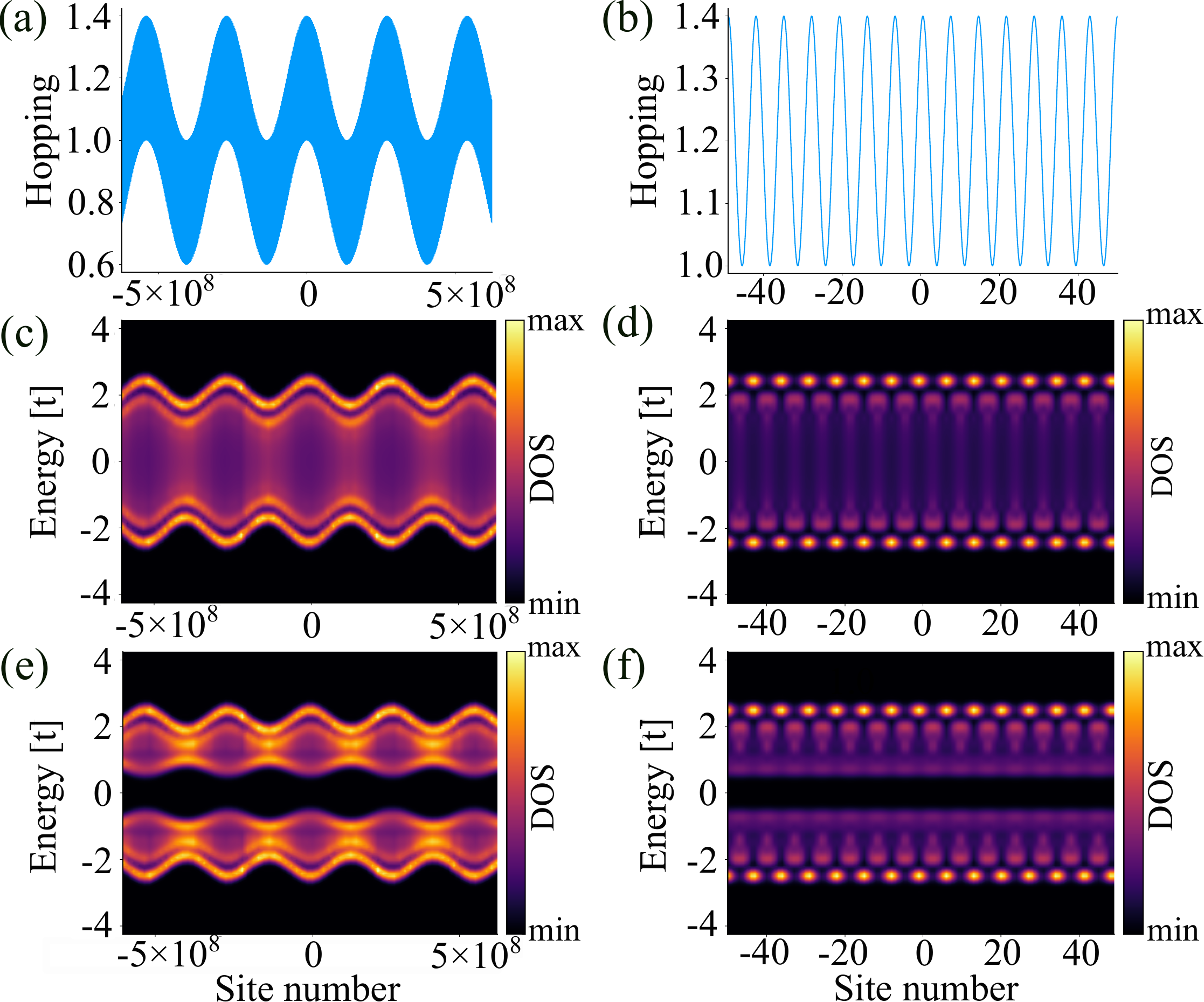}
    \caption{\textbf{One-dimensional super-moir\'e}. The super-moir\'e modulation is included in the hopping amplitudes. Panels (a) \& (b) show $t(X_{\alpha})$ at the larger moir\'e scale and smaller moir\'e scale. Panels (c) \& (d) show the corresponding local spectral functions calculated for a system with $2^{30}$ sites without Hubbard interaction at two different moir\'e scales. Panels (e) \& (f) show the corresponding local spectral functions calculated for a system with Hubbard interaction $U = 2.7$ at two different moir\'e scales.}
    \label{fig:hop}
\end{figure}

The previous Chebyshev expansion gives access to all the correlators
$\langle c^\dagger_{\alpha} c_{\beta} \rangle$.
The mean-field decoupling allows us to define a new mean-field Hamiltonian from
the tensor network representation of the correlators
combining Eq.~\ref{eq:cij}, Eq.~\ref{eq:kpm} and Eq.~\ref{eq:mf} as shown in Fig.~\ref{fig:1}(c).
Once the self-consistent mean-field Hamiltonian has been obtained, the
spatially dependent
spectral function $D(\omega,\alpha ) = 
\langle \alpha| \delta(\omega - \mathcal{H}^{MF}) |\alpha \rangle $ 
of the interacting problem can be computed
using a tensor network Chebyshev expansion as
\begin{equation}
D(\omega,\alpha ) = \langle \alpha| \left [
\sum_n \ T_n(\mathcal{H}^{MF}) P_n(\omega) 
\right ] | \alpha \rangle ,
\end{equation}
where $P_n (\omega) = \frac{2 T_n(\omega)}{\pi \sqrt{1-\omega^2}} $
for $n\ge1$ and $P_0 (\omega) = \frac{ T_0(\omega)}{\pi \sqrt{1-\omega^2}} $.

 \begin{figure}[t]
    \centering

    \includegraphics[width=1\linewidth]{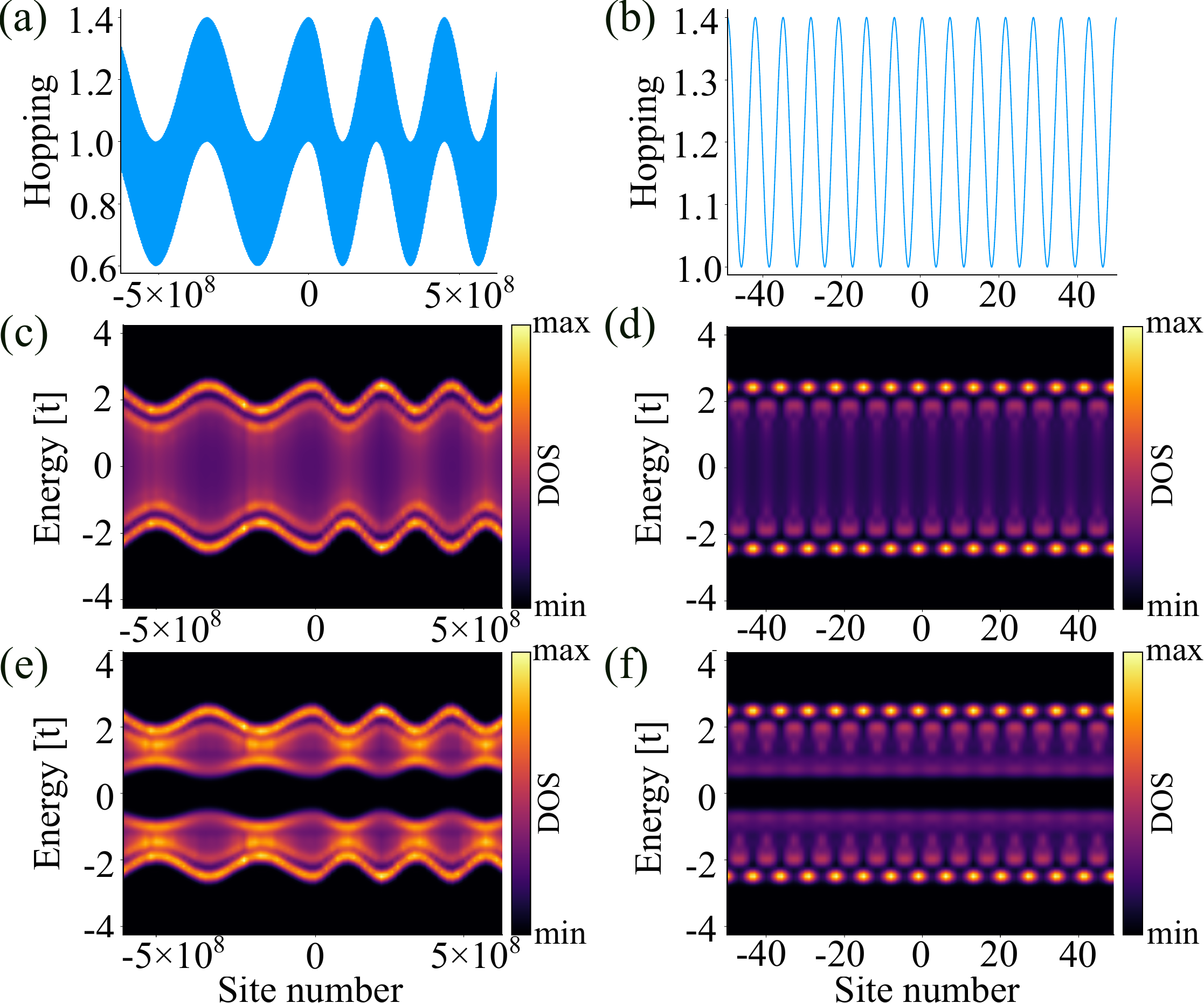}
    \caption{\textbf{Domain wall in a one-dimensional super-moir\'e}. The super-moir\'e modulation is included in the hopping amplitudes with a domain wall in the middle of the system. Panels (a) \& (b) show the hopping amplitudes $t(X_{\alpha})$ with the domain wall and two different larger moir\'e scales and smaller moir\'e scales. Panels (c) \& (d) show the corresponding local spectral functions calculated for a system with $2^{30}\gtrsim 10^9$ sites without Hubbard interaction at two different moir\'e scales. Panels (e) \& (f) show the corresponding local spectral functions calculated for a system with Hubbard interaction $U = 2.7$ at two different moir\'e scales.  }
    \label{fig:hop_domain}
\end{figure}

\section{Tensor network representation of the super-moir\'e mean-field Hamiltonians}
\label{construction}
We now elaborate on the explicit form of the MPO representation of the single-particle
Hamiltonian $H_0$ with more details provided in the Appendix. We first focus on the example of a one-dimensional spinful Fermi-Hubbard system, which has a Hamiltonian 
\begin{equation}
\begin{aligned}
    H &= \sum_{\alpha, s}t_{\alpha,\alpha+1}c_{x_{\alpha+1},s}^{\dagger}c_{x_{\alpha},s} + h.c.\\
    &+ \sum_{\alpha}U_\alpha 
    \left (c^\dagger_{x_{\alpha}, \uparrow} c_{x_{\alpha}, \uparrow} - \frac{1}{2}\right ) 
    \left (c^\dagger_{x_{\alpha}, \downarrow} c_{x_{\alpha}, \downarrow} - \frac{1}{2}\right ),
\end{aligned}
    \label{eq:HUbbard_half}
\end{equation}
where $t_{\alpha,\alpha+1}$ captures the spatially varying hopping amplitude between two adjacent atomic sites $x_{\alpha}$ and $x_{\alpha +1}$, and $U_\alpha$ is the Hubbard interaction induced by
the super-moir\'e correlation in a 1D moir\'e superlattice. 
We start with a 1D nearest-neighbor hopping $H_{0,NN} = \sum_{\alpha, s}^{N-1} t(c_{x_{\alpha+1},s}^{\dagger}c_{x_{\alpha},s} + h.c.)$.
The MPO representation of the uniform $\mathcal{H}_{0,NN}$ hopping term
could be built from~\cite{Jolly2025}
$
    \mathcal{H}_{0,NN} = \sum_{l,s}^{L} t(\sigma^{+}_{l,s}\bigotimes_{m>l} \sigma^{-}_{m,s} + h.c.),
$
where $\sigma^{\pm} = \frac{1}{2}(\sigma^{x} \pm i\sigma^{y})$, $\sigma^{x}$ and $\sigma^{y}$ being Pauli matrices. $l$ and $m$ are the site indices of the $L$-site MPO $\mathcal{H}_{0,NN}$.
We then address the case of a spatially modulated hopping amplitude $t_{\alpha,\alpha+1} = t(X_{\alpha})$, where $X_{\alpha} = \frac{x_{\alpha} + x_{\alpha+1}}{2}$.
Viewing $t(X_{\alpha})$ as a function of $X_{\alpha}$, by applying the quantics tensor cross interpolation  (QTCI)
algorithm~\cite{fernández2024learningtensornetworkstensor,Oseledets2010,Oseledets2011} over $t(X_{\alpha})$  
we now generate an
MPS representation $\ket{\tau} =  \sum M_{s_1}^{(1)} M_{s_2}^{(2)} M_{s_3}^{(3)}...M^{(L)}_{s_L}\ket{s_{1},s_{2},...,s_{L}}  $ storing all $t(X_{\alpha})$ values. For each $M_{s_j}^{(j)} $, we introduce an auxiliary index $s_{j}'$ to obtain a new tensor $\Gamma_{s_j,s_{j}'}^{(j)}$ and enforce a diagonal matrix structure of the data encoded with a contraction of $\Gamma_{s_j,s_{j}'}^{(j)}\delta_{j,j'}$. Repeating this process for all tensors in $\ket{\tau}$ yields an MPO representation $\mathcal{T}$ that stores a diagonal matrix whose elements are the same as those of $\ket{\tau}$. Eventually we perform a contraction between $\mathcal{T}$ and $\sum_{l,s}^{L}\sigma^{+}_{l,s}\bigotimes_{m>l} \sigma^{-}_{m,s}$, producing an MPO representation of the single-particle Hamiltonian
$
     \mathcal{H}_{0}   = \{[\mathcal{T} \sum_{l,s}^{L}   (\sigma^{+}_{l,s}\bigotimes_{m>l} \sigma^{-}_{m,s})] + h.c.\} .
$
The MPO $\mathcal{H}_{0}$ faithfully stores the original hopping terms, ensuring $\bra{x_{\alpha}, s}\mathcal{H}_{0}\ket{x_{\alpha+1},s}   \approx t(X_{\alpha})$.

\begin{figure}[t!]
    \centering
    \includegraphics[width=1\linewidth]{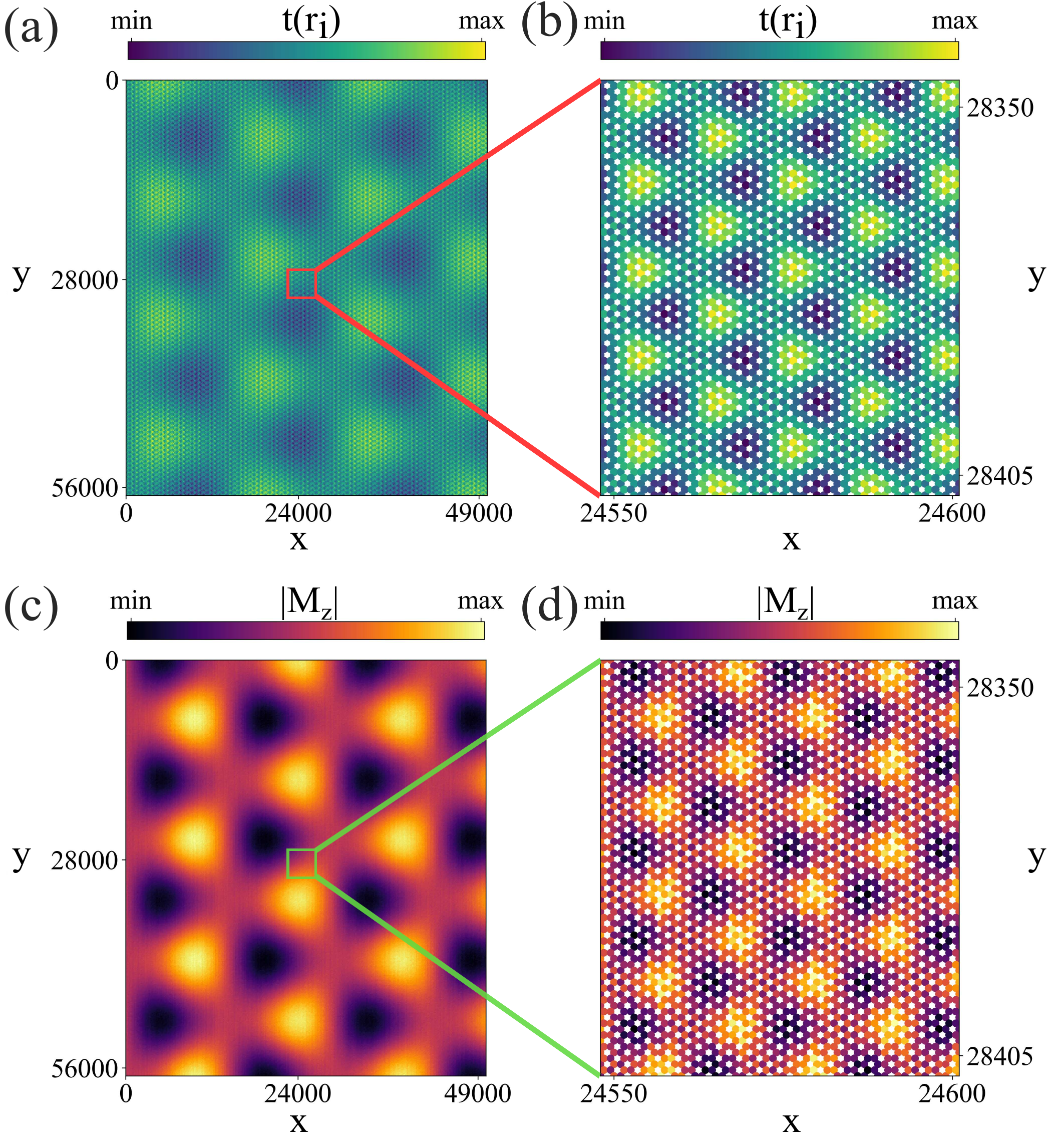}
    \caption{\textbf{Two-dimensional honeycomb super-moir\'e}. Results for $2^{31}$-site super-moir\'e modulated honeycomb lattice. Panels (a) and (b) shows the super-moir\'e modulations on the whole system and a center region of the system. Panels (c) and (d) show the magnetization distribution acquired from mean-field calculations corresponding to different length scales. }
    \label{fig::gra}
\end{figure}

\begin{figure}[t!]
    \centering
    \includegraphics[width=1\linewidth]{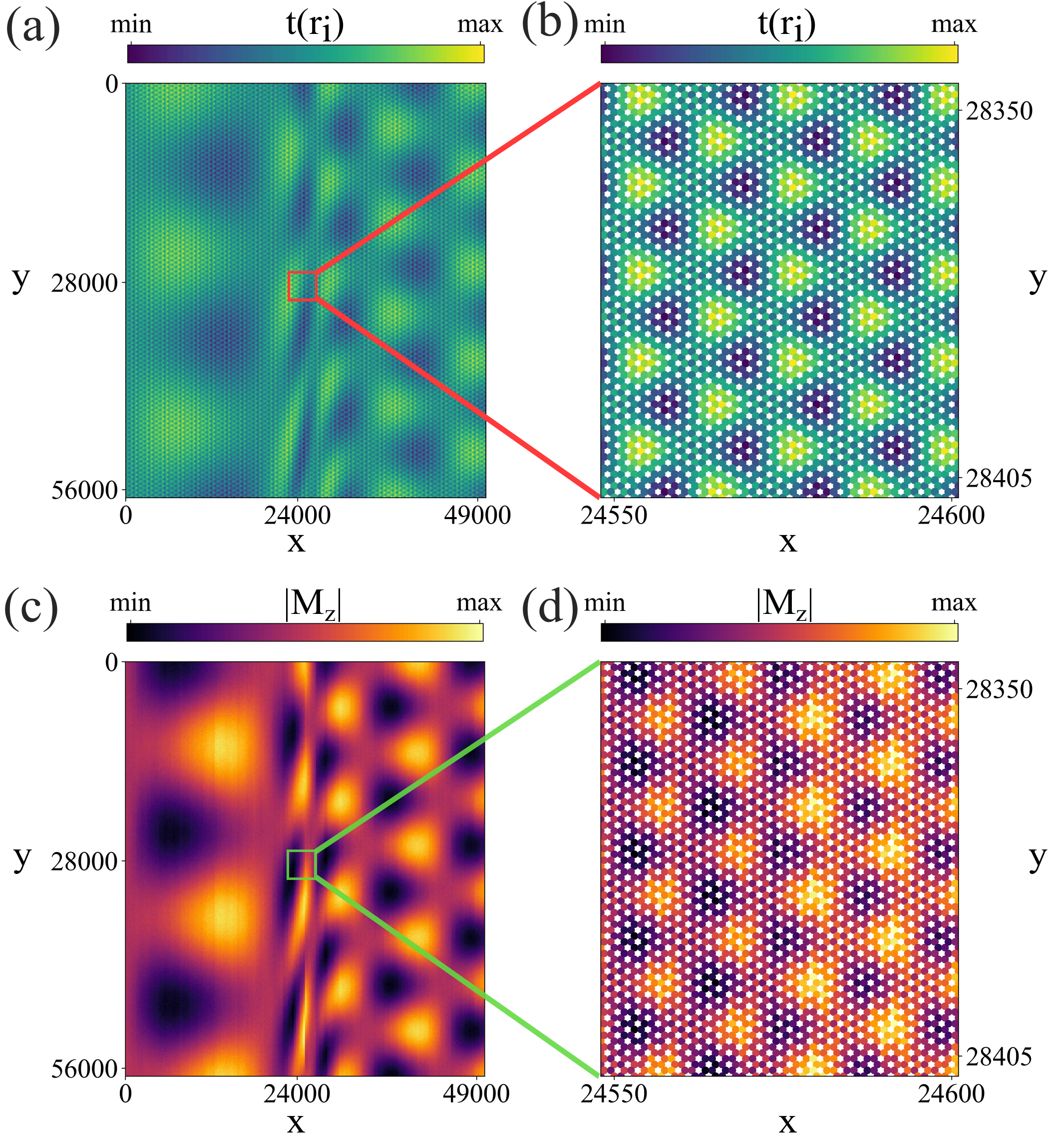}
    \caption{\textbf{Domain wall in a two-dimensional honeycomb super-moir\'e}. Results are for $2^{31}$-site super-moir\'e modulated honeycomb lattice with domain wall. Panels (a) and (b) show the super-moir\'e modulations with domain wall on the whole system and a center region of the system. Panels (c) and (d) show the magnetization distribution acquired from mean-field calculations corresponding to different length scales. }
    \label{fig::gra_domain}
\end{figure}

\section{Results}
\label{results}
\subsection{One-dimensional super-moir\'e system}
We now show results for the one-dimensional model of Eq.~\ref{eq:HUbbard_half}, 
with length of $2^{30} (\gtrsim 10^9)$ atomic sites. Such a model can be realized using optical quantum simulation methods~\cite{Fu2020,Wang2019} or in one-dimensional multi-walled nanotubes~\cite{caha2024magneticsinglewallcri3}. 
We assume $t(X_{\alpha})$ to have the form
$
t(X_{\alpha}) = 1 + 0.2\cos(k_{1}X_{\alpha}) + 0.2\cos(k_{2}X_{\alpha}),  
$
where we set $k_{1} = \frac{2\pi}{5 \sqrt{2}}$ and $k_{2} = \frac{8\pi}{2^{29}\sqrt{3}}$.   %They
These correspond to two different moir\'e scales as shown in Fig.~\ref{fig:hop}(a) and Fig.~\ref{fig:hop}(b), which are incommensurate with each other throughout the system.

The local spectral functions for systems with and without Hubbard interaction are shown in Figs.~\ref{fig:hop}(c-f).  
%The calculation for the self-consistent solution is shown in Fig.~\ref{fig:hop} (e) and (f).
This tensor network calculation featuring $2^{30}\gtrsim 10^9$ sites required
a computational time of $4\times10^4$ seconds on a single core of an Intel i5 CPU, leading to a speedup 
of around four orders of magnitude compared to the estimated runtime that would be required by a previous sparse-matrix-based approach~\cite{Fumega2024}, which is in practice infeasible for the system sizes considered here.
The interacting solution shows modulations of the spectral function
following the larger moir\'e scale corresponding to $k_{2}$ and the smaller moir\'e scale corresponding to $k_{1}$. In the presence of Hubbard on-site interactions, a spatially inhomogeneous gap emerges across the system, which exhibits spatial variations governed by the larger moir\'e scale yet remains nearly uniform at the smaller moir\'e scale. This is due to the fact that the scale of the Hubbard interaction considered in this setup is of the same order as the smaller moir\'e scale, while being negligible in comparison to the larger moir\'e scale. Consequently, a uniform gap opens across all sites at the smaller moir\'e scale while being explicitly modulated by the super-moir\'e correlation at the larger moir\'e scale.

Next we use the same form of Hamiltonian as in Eq.~\ref{eq:HUbbard_half} but include a domain wall by modulating the larger moir\'e scale $ k_{2}$ as
$
   \tilde{k}_{2} = k_{2}[1 + \delta\tanh(\frac{X_{\alpha}}{W})],    
$
where $\delta$ is a parameter used to quantify the mismatch between the two %2
larger moir\'e scales separated by the domain wall, and $W$ is the width of the domain wall. We set $\delta = 0.2$, $W = \frac{2^{30}}{40}$, and keep other parameters consistent with the calculation without the presence of the domain wall. In Figs.~\ref{fig:hop_domain}(c, e), it is clear that the local spectral functions closely follow the modulation of larger moir\'e scale on both sides of the domain wall. The opening of the gap due to the Hubbard interaction shown in Fig.~\ref{fig:hop_domain}(e) also obeys this modulation, while at the smaller moir\'e scale, spectral functions are not influenced by the domain wall as shown in Figs.~\ref{fig:hop_domain}(d, f).

\subsection{Two-dimensional super-moir\'e system}

To demonstrate the capability of our methodology in handling two-dimensional super-moir\'e systems, we first conduct mean-field calculations for super-moir\'e graphene, a platform that has shown great potential in recent experiments~\cite{Park2025, Ma2025}. 
We account for the super-moir\'e modulation of hopping amplitudes in systems both with and without domain walls. In all cases, the Hubbard interaction is fixed at $U = 5.5t$.
We consider systems with $2^{31}$ sites which have the shape of $2^{16}$ sites times $2^{15}$ sites for better visualization of the results. 
For the case with super-moir\'e modulation, the hopping amplitudes take the form 
\begin{equation}
    t_{ij} = 1 + 0.2\sin(k_1\mathbf{u}_{ij}\cdot \mathbf{R}_{ij}) + 0.2\sin(k_2\mathbf{u}_{ij}\cdot \mathbf{R}_{ij}),
    \label{eq:supmoir}
\end{equation}
with $k_1 = \frac{\pi}{2^{12}\sqrt{3}}$ and $k_2 = \frac{\pi}{4\sqrt{2}}$,
$\mathbf{u}_{ij}$ the vector linking site $i$ and $j$,
and $\mathbf{R}_{ij} = \frac{\mathbf{r}_i+\mathbf{r_j}}{2}$, $\mathbf{r}_{i}$ the average of
positions $i$ and $j$. Fig.~\ref{fig::gra} (a) and (b) display the modulations on all sites of the systems as $t(r_i) = \sum_{j}t_{ij}$.

Then we consider the existence of a domain wall in the $x$ direction of the system as shown in Fig.~\ref{fig::gra_domain} (a) and (b). In this case we  take the hopping amplitudes as 
\begin{equation}
    t_{ij} = 1 + 0.2\sin(\tilde{k}_1\mathbf{u}_{ij}\cdot \mathbf{R}_{ij}) + 0.2\sin(k_2\mathbf{u}_{ij}\cdot \mathbf{R}_{ij}),
    \label{eq:domain}
\end{equation}
while $\tilde{k}_{1}$ only depends on $X_{ij} = \frac{x_i + x_j}{2}$  
\begin{equation}
    \tilde{k}_{1} = k_1(1+\delta\tanh(\frac{X_{ij}}{W})),
\end{equation}
here $\delta = 0.3$ and $W = 3000$. In Fig.~\ref{fig::gra} and Fig.~\ref{fig::gra_domain}, we observe that the magnetization $M_z = |\rho_{\uparrow} - \rho_{\downarrow}|$, where $\rho$ is the spin-resolved electron density, accurately reflects the hopping modulations across all scales. 
This agreement confirms the robustness of our method in resolving the real-space structure of super-moir\'e patterns in ultra-large systems,
and it is able to simultaneously capture super-moir\'e and domain walls.

\begin{figure}[t!]
    \centering
 
    \includegraphics[width=1\linewidth]{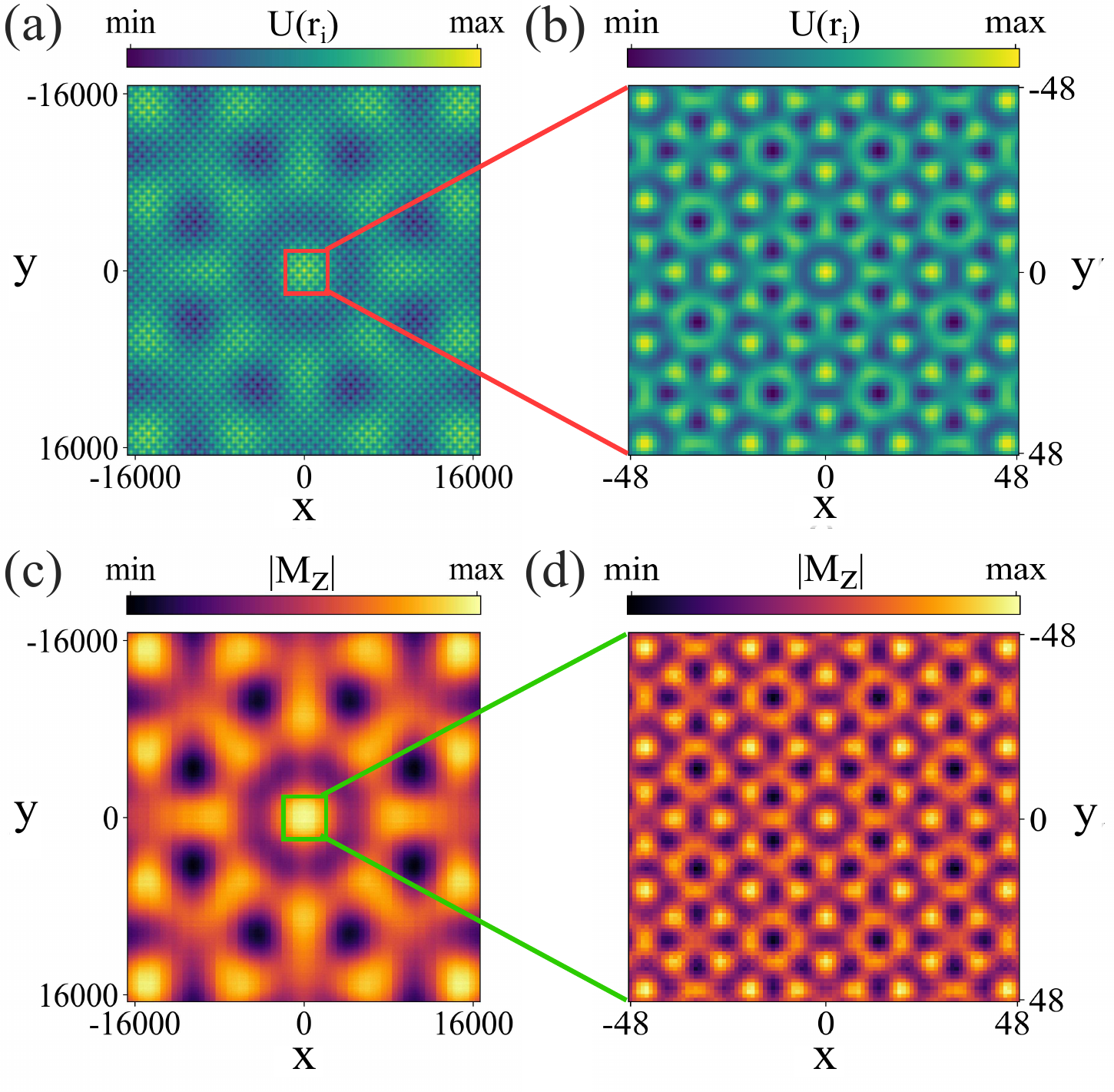}
    \caption{\textbf{Two-dimensional eightfold square super-moir\'e quasi-crystal}. Results for $2^{30}$-site square lattice with super-moir\'e Hubbard interaction with eightfold rotational symmetry. Panels (a) \& (b) show the Hubbard interaction $U(r_i)$ at the larger moir\'e scale and smaller moir\'e scale. Panels (c) \& (d) show the corresponding magnetization $|M_z|$ calculated for the system at two different moir\'e scales. }
    \label{fig:inter}
\end{figure}

Furthermore, our methodology can be applied to systems featuring quasi-crystalline structures, and when the modulations
are not in the hopping but the electronic interactions.
We demonstrate this in a square lattice with a modulated
Hubbard interaction with $2^{30}$ ($2^{15}\times2^{15}$) sites. Such systems can be realized through Coulomb engineering~\cite{PhysRevResearch.3.013265,Jiang2021,PhysRevLett.133.206601}. 
To demonstrate the capability of our method in handling more complex spatial structures, we introduce a 
modulation on the Hubbard interaction with approximate eightfold rotational symmetry for simulating a quasi-crystalline system~\cite{Uri2023, Lai2025, Hao2024} as  
$
    U(r_i) = 8 + 0.5\{\sum_{i=1}^{4}\cos(k_{1}\mathbf{n}_{i}\cdot \mathbf{r}_i) + \cos(k_{2}\mathbf{n}_{i}\cdot \mathbf{r}_i) \},
$
where the set of unit vectors ${\mathbf{n}_j}$ includes $(1, 0)^T$, $(0, 1)^T$, $(\frac{\sqrt{2}}{2}, \frac{\sqrt{2}}{2})^T$, and $(-\frac{\sqrt{2}}{2}, \frac{\sqrt{2}}{2})^T$, corresponding to four directions that collectively span an approximate eightfold symmetric basis in two dimensions and $k_1 = \frac{\pi}{2^{11}\sqrt{3}}$, $k_2 = \frac{\pi}{2 \sqrt{5}}$. As shown in Figs.~\ref{fig:inter}(a-d), the spatial distribution of local magnetization closely follows the modulation of $U(r_i)$ at both large and small scales. These results confirm that our method accurately captures spatially modulated correlated states and is readily applicable to a wide range of realistic super-moir\'e and quasi-crystalline systems.

 A key advantage of our approach is that it bypasses the need to store the single-particle Hamiltonian as sparse matrices like in prior methods~\cite{Fumega2024}. This enables simulations of systems over three orders of magnitude larger. Representing the full Hamiltonian as an MPO also allows KPM calculations without evaluating Chebyshev moments with separate basis vectors, which significantly boosts efficiency. However, this advantage relies on the structured nature of the Hamiltonian in super-moir\'e systems, where the modulations can be expressed as deterministic functions of spatial coordinates. In contrast, strong randomness in highly disordered systems will significantly hinder the compression of system Hamiltonian via the QTCI algorithm.
Furthermore, by encoding the mean-field Hamiltonian as MPOs, our method naturally interfaces with established tensor network algorithms such as real-time evolution~\cite{PhysRevLett.107.070601,Paeckel2019,niedermeierGP,boucomasGP,chenGP} and can therefore be extended to compute dynamics of super-moir\'e materials in real space. While our discussion has focused on tight-binding models,
it is worth noting that our tensor network mean-field
methodology can be readily extended to real space
density functional theory~\cite{thijssen2007computational, pisani2012hartree,Enkovaara2010,2025arXiv250308430H},
which can potentially enable an increase in the system sizes that these methods can reach.

\section{Conclusion} 
\label{conclusion}
In summary, we have established a methodology that allows us to solve billion-size interacting super-moir\'e problems
by using a self-consistent tensor network algorithm. 
Our strategy uses a mapping from
the real space Hamiltonian of a system with a size of $2^{L}$ atomic sites into a
tensor network representation of a many-body system corresponding to a chain of $L$-site pseudo spins. 
By using a combination of tensor network Chebyshev expansion and quantics tensor
cross interpolation,
we can solve the self-consistent mean-field equations
of ultra-large systems. 
We demonstrate this technique with super-moir\'e interacting models above a billion sites,
featuring a spatially dependent single-particle term and spatially dependent many-body interactions,
including domain walls.
Our work establishes a widely applicable technique to treat interacting
electronic problems, %allowing
providing a method to solve correlated systems of unprecedented sizes, and in particular
whose single-particle Hamiltonian alone is too large to be explicitly stored even
with sparse representations.
Ultimately, our work provides a starting point
to perform many-body perturbation theory 
and linear response
using tensor networks in real space,
providing the required method to rationalize a wide range
of correlated phenomena in super-moir\'e quantum matter.

\textbf{Acknowledgments:}
We acknowledge the computational resources provided by the Aalto Science-IT project and the financial support from InstituteQ,  
the Jane and Aatos Erkko Foundation, the Academy of Finland Projects Nos. 331342, 358088, 349696, and 369367 Finnish Ministry of Education and Culture through the Quantum Doctoral Education Pilot Program and the Research Council of Finland through the Finnish Quantum Flagship project No. 358878,
and the ERC Consolidator Grant ULTRATWISTROICS (Grant agreement no. 101170477).  
We thank X. Waintal, C. Flindt, C. Yu, P. San-Jose, E. Prada, B. Amorim, E. Castro, P. Alcazar, M. Nguyen and R. Oliveira for useful discussions. The code used in this work is available at~\cite{myrepo}.

\appendix

\section{Construction of matrix product operator Hamiltonians for one-dimensional systems}

In the following we elaborate on how we construct the MPOs for the hopping terms used in our 1D and 2D mean-field Hamiltonians. For simplicity we neglect the electron spin. starting with the interacting term, assume the Hubbard interaction is a function defined throughout the system as $U(x_i)$, one can directly construct a diagonal MPO $\mathcal{U}$ storing all the function values using QTCI algorithm~\cite{Jolly2025} with matrix form as
\begin{equation}
            \begin{bmatrix}
            U(x_1) & 0 & 0 & 0 & \cdots & 0 \\
            0 & U(x_2) & 0 & 0 & \cdots & 0 \\
            0 & 0 & U(x_3) & 0 & \cdots & 0 \\
            \vdots & \vdots & \vdots & \ddots & \vdots & \vdots \\
            0 & 0 & 0 & \cdots & U(x_{2^L-1}) & 0 \\
            0 & 0 & 0 & \cdots & 0 & U(x_{2^L})  
        \end{bmatrix},
\end{equation}
where $x_i$ is the spatial position of site $i$. Then with the electron density diagonal MPO $\mathcal{N}$ having the matrix form of
\begin{equation}
            \begin{bmatrix}
            n(x_1) & 0 & 0 & 0 & \cdots & 0 \\
            0 & n(x_2) & 0 & 0 & \cdots & 0 \\
            0 & 0 & n(x_3) & 0 & \cdots & 0 \\
            \vdots & \vdots & \vdots & \ddots & \vdots & \vdots \\
            0 & 0 & 0 & \cdots & n(x_{2^L-1}) & 0 \\
            0 & 0 & 0 & \cdots & 0 & n(x_{2^L})  
        \end{bmatrix},
\end{equation}
which is generated in each self-consistent field (SCF) iteration with QTCI algorithm, the correct on-site interaction term is the contraction $\mathcal{U}\mathcal{N}$.

Next, we start with the fundamental ingredient of our construction of all types of hopping Hamiltonian. For the L-site MPO of $\mathcal{S} = \sum_{l}^{L}\sigma^{+}_{l}\bigotimes_{m>l} \sigma^{-}_{m}$, the matrix form of $S$ will be
\begin{equation}
    \begin{bmatrix}
0 & 1 & 0 & 0 & \cdots & 0 \\
0 & 0 & 1 & 0 & \cdots & 0 \\
0 & 0 & 0 & 1 & \cdots & 0 \\
0 & 0 & 0 & 0 & \ddots & 0 \\
\vdots & \vdots & \vdots & \vdots & \ddots & 1 \\
0 & 0 & 0 & 0 & \cdots & 0
\end{bmatrix},
\end{equation}
with the shape of $2^L \times 2^L$. Then if the hopping amplitudes of this system are position-dependent as shown in the main text which could be written as a function of spatial coordinate as $t(X)$ where $X_{i} = \frac{x_{i} + x_{i+1}}{2}$, one can follow the approach in the paper using QTCI algorithm and construct a diagonal MPO $\mathcal{T}$ whose matrix form will be
\begin{equation}
            \begin{bmatrix}
            t(X_1) & 0 & 0 & 0 & \cdots & 0 \\
            0 & t(X_2) & 0 & 0 & \cdots & 0 \\
            0 & 0 & t(X_3) & 0 & \cdots & 0 \\
            \vdots & \vdots & \vdots & \ddots & \vdots & \vdots \\
            0 & 0 & 0 & \cdots & t(X_{2^L - 1}) & 0 \\
            0 & 0 & 0 & \cdots & 0 & t(X_{2^L})  
        \end{bmatrix},
\end{equation}
For simplicity, here we assume all functions we have for hopping amplitudes are real. Then the contraction between $\mathcal{T}$ and $\mathcal{S}$ will result in the matrix
\begin{equation}
\begin{aligned} 
    \begin{bmatrix}
        0 & t(X_1) & 0 & 0 & \cdots & 0 & 0 \\
        0 & 0 & t(X_2) & 0 & \cdots & 0 & 0 \\
        0 & 0 & 0 & t(X_3) & \cdots & 0 & 0 \\
        0 & 0 & 0 & 0 & \ddots & \vdots & \vdots \\
        0 & 0 & 0 & 0 & \ddots & t(X_{2^L - 2}) & 0 \\
        0 & 0 & 0 & 0 & 0 & 0 & t(X_{2^L - 1}) \\
        0 & 0 & 0 & 0 & 0 & 0 & 0
    \end{bmatrix}.
\end{aligned}
\end{equation}
In such a contraction, the $t(X_{2^L}) $ term is discarded and all other hopping amplitudes are correctly cast onto $\mathcal{S}$. To construct the other half of the Hamiltonian, one only needs to construct the hermitian conjugate $(\mathcal{T}\mathcal{S})^{\dag}$ as
\begin{equation}
\begin{aligned} 
    \begin{bmatrix}
        0 & 0 & 0 & 0 & \cdots & 0 & 0 \\
        t(X_1) & 0 & 0 & 0 & \cdots & 0 & 0 \\
        0 & t(X_2) & 0 & 0 & \cdots & 0 & 0 \\
        0 & 0 & t(X_3) & 0 & \cdots & 0 & 0 \\
        \vdots & \ddots & \ddots & \ddots & \ddots & \vdots & \vdots \\
        0 & \cdots & \cdots & \cdots & t(X_{2^L - 2}) & 0 & 0 \\
        0 & \cdots & \cdots & \cdots & 0 & t(X_{2^L - 1}) & 0
    \end{bmatrix}  .  
\end{aligned}
\end{equation}
Summing the two components $\mathcal{T}\mathcal{S}+ (\mathcal{T}\mathcal{S})^{\dag}$ we are able to construct the hopping MPO for one-dimensional nearest-neighbor hopping systems. Then, $\mathcal{T}\mathcal{S}+ (\mathcal{T}\mathcal{S})^{\dag} + \mathcal{U}\mathcal{N}$ will result in the full Hamiltonian expressed in the form of MPO.

It is straightforward to extend this approach to long-range hoppings. For the construction of next-nearest-neighbor hopping MPO, it will simply be the square of $\mathcal{S}$, as the matrix form of such square of $\mathcal{S}$ is
\begin{equation}
    \begin{bmatrix}
0 & 0 & 1 & 0 & \cdots & 0 \\
0 & 0 & 0 & 1 & \cdots & 0 \\
0 & 0 & 0 & 0 & \cdots & 0 \\
0 & 0 & 0 & 0 & \ddots & 1 \\
\vdots & \vdots & \vdots & \vdots & \ddots & 0 \\
0 & 0 & 0 & 0 & \cdots & 0
\end{bmatrix},    
\end{equation}
which shifts the values on first off-diagonal line to second off-diagonal line. Through this approach, we can construct MPOs for shifting matrix elements for even larger distances. For the notations, we call the original $\mathcal{S}$ to be $\mathcal{S}_{0}$ as it will shift the matrix elements of any MPO it applies on by $2^0 = 1$ line. Such shifting MPOs allow one to easily construct hopping MPOs for large-scale systems with arbitrary long-range hoppings or couplings. For example, to construct the shifting MPO $\mathcal{S}_{2^N}$ where $N$ is an arbitrary integer, one simply needs to do
\begin{equation}
    \mathcal{S}_{2^N} = \prod_{i = 1}^N \mathcal{S}_{1} = \prod_{i = 1}^N (\mathcal{S}_{0})^2 .
\end{equation} 
This example shown here will be significant for the construction of hopping MPOs for two-dimensional systems. On the other hand, to shift the lines down, one can use the hermitian conjugate of these shifting MPOs. Eventually since every decimal number can be expressed with a binary expression, in principle our approach allows one to shift the diagonal elements to arbitrary positions in the matrix, making it possible to construct the MPO for Hamiltonians with complex structures.

\begin{figure}[t!]
    \centering
    \includegraphics[width=1\linewidth]{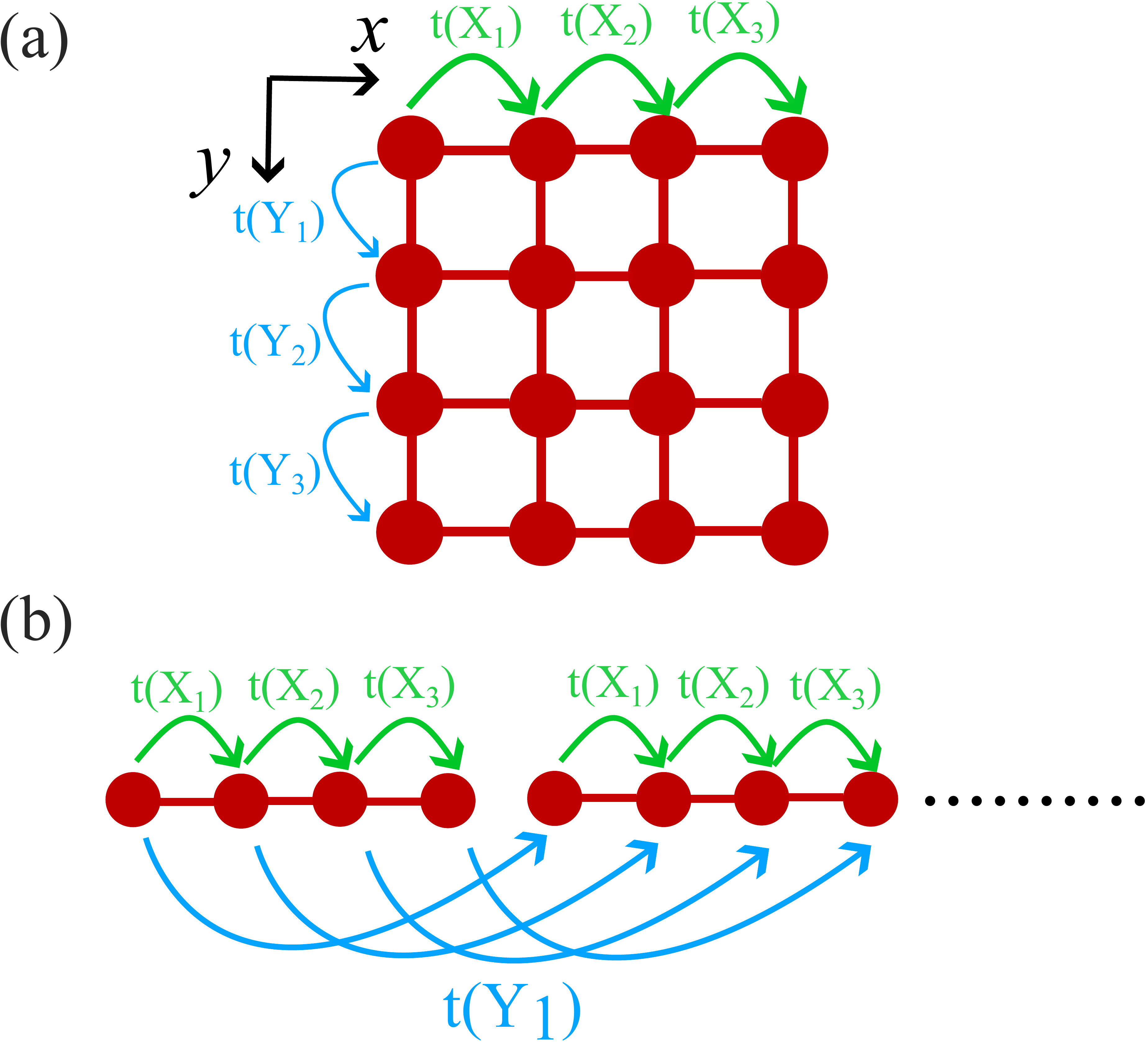}
    \caption{\textbf{Schematic of the construction of two-dimensional Hamiltonians}. A 16-site square lattice example to show how we map the two-dimensional system to one-dimensional chains with intra- and inter-chain hoppings. Panel (a) shows the original system where we mark the hopping amplitudes in $x$ direction with $t(X_{i})$ and in $y$ direction $t(Y_{i})$. Panel (b) demonstrates the equivalent 1D chains showing that the original $t(X_{i})$ become intra-chain hoppings and $t(Y_{i})$ the interchain hoppings.}
    \label{fig:example}
\end{figure}

\section{Construction of matrix product operator for two-dimensional systems}

Next we explain how to construct hopping MPOs for two-dimensional systems. The core idea is to view the 2D lattices as separate 1D chains with intra-chain and inter-chain hoppings as shown in Fig.~\ref{fig:example}.  

We can first consider a square lattice with size of $2^M \times 2^M$. If we assume that intra-chain hopping has the direction of $x$ and inter-chain hopping has the direction of $y$, we can construct 2 diagonal MPOs $\mathcal{T}_{intra}$ and $\mathcal{T}_{inter}$ with $2M$ sites based on $X_{i}$ and $Y_{i} = \frac{y_i+y_{i+1}}{2}$. The $\mathcal{T}_{intra}$ should have the form of $\bigoplus^{2^M}\mathcal{T}_{X}$, where $\mathcal{T}_{X}$ has the matrix form of
\begin{equation}
\begin{bmatrix}
    t(X_1) & 0 & 0 & 0 & \cdots & 0 \\
    0 & t(X_2) & 0 & 0 & \cdots & 0 \\
    0 & 0 & t(X_3) & 0 & \cdots & 0 \\
    \vdots & \vdots & \vdots & \ddots & \vdots & \vdots \\
    0 & 0 & 0 & \cdots & t(X_{2^M - 1}) & 0 \\
    0 & 0 & 0 & \cdots & 0 & t(X_{2^M})  
\end{bmatrix},
\end{equation}
so that this block repeats $2^M$ times for all the chains in the system. While $\mathcal{T}_{inter}$ should have the matrix form of $\bigoplus_{j=1}^{2^M} \mathcal{T}_{Y_{j}}$ where $\mathcal{T}_{Y_{j}} = t(Y_j)(\mathcal{I}_{2^M})$,   $\mathcal{I}_{2^M}$ being the MPO of identity matrix with shape of $2^M \times 2^M$. In this way, all the hopping amplitudes between 2 neighboring chains are the same, as shown in Fig.~\ref{fig:example}(b).

We can then construct the correct hopping MPO using $\mathcal{T}_{intra}$ and $\mathcal{T}_{inter}$. First of all, the intra-chain hoppings will be very similar as in the 1D cases. One just needs to notice that from the site with index of $2^M$ to the site with index $2^M + 1$, there is no actual hopping in the 2D cases since this is where the first chain ends and the second chain starts, which is shown in Fig.~\ref{fig:example}(b). Thus after the generation of the diagonal MPO $\mathcal{T}_{intra}$, an additional operator $\mathcal{B}$ will be needed to erase the diagonal values exactly at positions proportional to the length of the chain $2^M$. The matrix form of it will be   
\begin{equation}
\bigoplus^{2^M} 
\begin{bmatrix}
\mathbf{I}_{2^M - 1} & \mathbf{0}_{1\times(2^M-1)} \\
\mathbf{0}_{ (2^M-1) \times 1} & 0
\end{bmatrix}.
\end{equation}
This MPO could be constructed either from the QTCI algorithm or one can construct it in a pure product of tensor approach as
\begin{equation}
    \mathcal{B} = \bigotimes^{2M}_{i=1} \sigma_{i}^{0}- (\bigotimes^{2M}_{i=1}\sigma_{i}^{0})\prod_{j=M+1}^{2M}\sigma^{d}_{j}, 
\end{equation}
where $\sigma_{i}^{0}$ is the identity matrix operator acting on the $i$th site of tensor network and $\sigma_{i}^{d} = \frac{1}{2}(\sigma_{i}^{0}-\sigma_{i}^{z} )$. Then the correct intra-chain hopping MPO will be $\mathcal{B}\mathcal{T}_{intra}\mathcal{S}_0 + h.c.$, where $\mathcal{B}\mathcal{T}_{intra}\mathcal{S}_0$ will have the matrix form of
\begin{equation}
\left(
\bigoplus^{2^M - 1}
\left(
T_{shifted} \oplus 0
\right)
\right)
\oplus T_{shifted} ,
\end{equation}
and the $T_{shifted}$ is the matrix of
\begin{equation}
\begin{bmatrix}
0 & t(X_1 ) & 0 & \cdots & 0 & 0 \\
0 & 0 & t(X_2 ) & \cdots & 0 & 0 \\
\vdots & \vdots & \ddots & \ddots & \vdots & \vdots \\
0 & 0 & \cdots & 0 & t(X_{2^M-2} ) & 0 \\
0 & 0 & \cdots & 0 & 0 & t(X_{2^M-1} ) \\
0 & 0 & \cdots & 0 & 0 & 0 \\
\end{bmatrix}.
\end{equation}

For the inter-chain hopping terms, once the diagonal MPO $\mathcal{T}_{inter}$ is constructed, it can be correctly positioned by shifting it upward by $2^M$ lines. As illustrated in Fig.~\ref{fig:example}, the site index difference between two coupled sites in adjacent chains is exactly $2^M$, meaning the corresponding off-diagonal terms in the original Hamiltonian appear $2^M$ lines above the main diagonal. Therefore, shifting $\mathcal{T}_{inter}$ upward by $2^M$ aligns it with the correct off-diagonal structure in the full Hamiltonian. As discussed in previous section, this only requires a shifting operator $\mathcal{S}_{2^M}$ and could simply be finished with $M$ times of contraction of $\mathcal{S}_1$. Then the correct MPO for the inter-chain hopping for square lattice will simply be $\mathcal{T}_{inter}\mathcal{S}_{2^M}+h.c.$. Eventually the correct MPO for the the hopping Hamiltonian of square lattice is $\mathcal{B}\mathcal{T}_{intra}\mathcal{S}_0 + \mathcal{T}_{inter}\mathcal{S}_{2^M} + h.c.$

Furthermore, starting from the construction of the hopping MPOs for the square lattice, one can readily extend the approach to triangular and honeycomb lattices with only few modifications. The construction of intra-chain hopping MPOs remains identical to that of the square lattice. While the inter-chain hopping MPOs are more intricate in these lattices, the same underlying strategy applies: construct the appropriate diagonal MPOs for the hopping terms, shift each component to its correct position, and finally assemble the full Hamiltonian in MPO form.

\section{General Comments on Spectral Rescaling and Hamiltonian Compressibility}

Following our explanations for constructing MPOs, we provide here additional discussion to contextualize the scope and robustness of our methodology. Since our approach combines several advanced numerical techniques, we focus on two central aspects: the application of the kernel polynomial method (KPM) over MPOs, and the compressibility of the Hamiltonian via the QTCI algorithm.

\subsection{Spectral Rescaling of the Mean-Field Hamiltonian}

Performing the calculations with the KPM algorithm require a spectral rescaling of the Hamiltonian of the system~\cite{RevModPhys.78.275}. In several previous works where KPM method is applied to MPO-form Hamiltonians~\cite{Yang2020, Holzner2011, PhysRevLett.130.100401,PhysRevResearch.1.033009}, careful considerations must be paid to the rescaling of the Hamiltonians onto the energy window of $(-1,1)$. However, a key distinction between our work and prior studies lies in the type of Hamiltonian considered. While the cited references typically focus on quantum many-body Hamiltonians, our method operates on single-electron mean-field Hamiltonians. In effectively
single-particle Hamiltonian, as is our mean-field Hamiltonian in MPO form, 
the bandwidth is size independent even in the presence of mean-field interactions. 
Specifically, for a one-dimensional modulated system, the band width of the MPO Hamiltonian
is approximately $W \approx 4t + U$, whereas for a two-dimensional square or honeycomb lattice
it is approximately $W \approx 8t + U$ or $W \approx 6t + U$. As a result, in our manuscript spectral functions do
not have long tails. This key difference greatly simplifies the rescaling step and mitigates the risk of pathological spectral behavior as the system size increases.

In our implementation, we normalize the single-particle Hamiltonian by an overall factor to ensure that its spectrum falls within the $(-1,1)$ interval required for the Chebyshev expansion. This simple procedure has proven robust in all models considered, including large-scale 1D and 2D super-moir\'e systems. If the bandwidth is not known in advance, it can be estimated via density matrix renormalization group method applied to the MPO to obtain the extremal eigenvalues, which define the required scaling window.

\begin{figure*}[t!]
    \centering
    \includegraphics[width=1\linewidth]{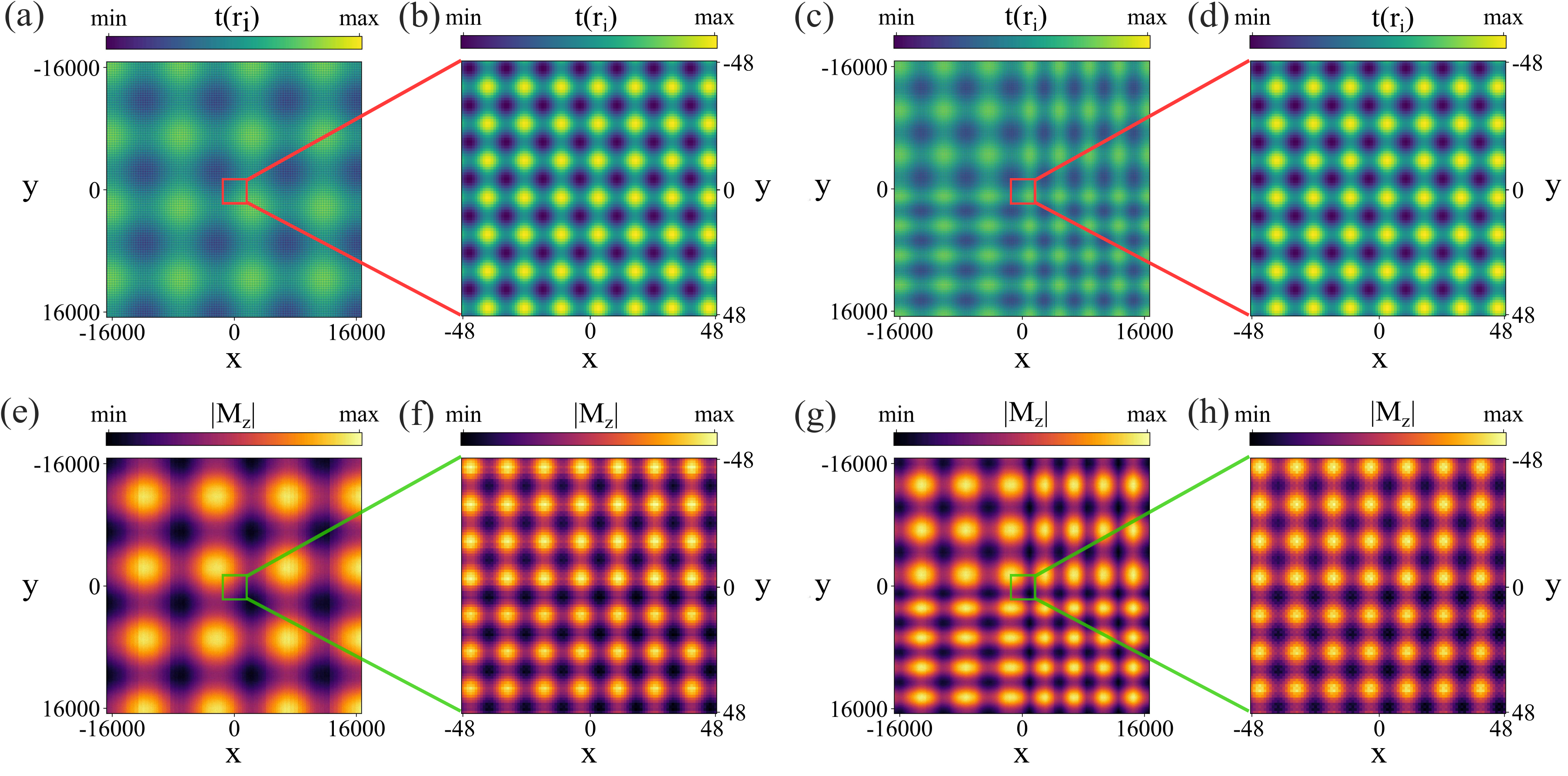}
    \caption{\textbf{Two-dimensional square super-moir\'e lattice}. Results for $2^{30}$-site super-moir\'e modulated square lattice with and without domain walls. Panels (a) and (b) shows the super-moir\'e modulations on the whole system and a center region of the system. Panels (c) and (d) shows the super-moir\'e modulations with domain walls on the whole system and a center region of the system. Panels (e-h) show the magnetization distribution acquired from mean-field calculations corresponding to different systems and length scales following Panels (a-d).}
    \label{fig:square}
\end{figure*}

\subsection{Compressibility of the Hamiltonian}

In the main text we comment about the "structured" Hamiltonian of super-moir\'e system makes it possible for the compression of them as MPOs. The notion of “structured” in our context refers to the property that the modulation in the Hamiltonian (e.g., moir\'e or super-moir\'e patterns) can be expressed as a deterministic function of spatial coordinates, even if the system lacks translational invariance. This functional structure is what allows the application of the QTCI algorithm to efficiently build a tensor train storing all function values for further MPO construction. The capability of QTCI algorithm in approximating physical observables featuring complicated functions has already been proved in various former works~\cite{10.21468/SciPostPhys.18.1.007,fernández2024learningtensornetworkstensor,Jolly2025, PhysRevX.12.041018}. Specifically, our algorithm relies on the compressibility of the Hamiltonian, which enables finding a tensor network representation with small bond dimension. Quasi-periodic and quasi-crystalline systems fall squarely into this category and are naturally compressible. 

For highly disordered systems where every single site is independent from every other one, our methodology would not be effective as such Hamiltonian could not be compressed as a tensor network with small bond dimension. In comparison, sparse disorder, including dilute impurities, twist disorder
or domain walls are compressible, as the profiles for those systems can be deterministically stored and generated with small resources~\cite{fernández2024learningtensornetworkstensor}. Specifically, the 1D and 2D domain wall results presented on our revised manuscript belong
to this sparse disorder scenario.
From the perspective
of tensor networks, compressibility is associated to a matrix product state with a low bond dimension, which represents modulations where the entanglement between length scales is not
excessively large~\cite{fernández2024learningtensornetworkstensor}.

\section{Additional results for two-dimensional systems}

Beyond the two-dimensional results shown in Fig. \ref{fig::gra}, Fig. \ref{fig::gra_domain} and Fig. \ref{fig:inter} of the main text, we show here analogous results for a super-moir\'e two-dimensional square lattice. 
We consider the super-moir\'e modulation of the hopping amplitudes and for systems with and without domain walls, 
where the Hubbard interaction is set to be a constant of $U = 5.5t$.

First we consider super-moir\'e square lattice with $2^{30}$ sites ($2^{15}\times2^{15}$). Such system could be simulated with photonic lattices and twisted materials~\cite{Zhang2022, https://doi.org/10.48550/arxiv.2406.05626}. The hopping amplitudes of the system takes the form
\begin{equation}
    t_{ij} = 1 + 0.2\sin(k_1\mathbf{u}_{ij}\cdot \mathbf{R}_{ij}) + 0.2\sin(k_2\mathbf{u}_{ij}\cdot \mathbf{R}_{ij}),
    \label{eq:supmoir}
\end{equation}
where $k_1 = \frac{\pi}{2^{11}\sqrt{5}}$ and $k_2 = \frac{\pi}{4\sqrt{3}}$. $\mathbf{u}_{ij}$ is the vector linking site $i$ and $j$ while $\mathbf{R}_{ij} = \frac{\mathbf{r}_i+\mathbf{r_j}}{2}$, $\mathbf{r}_{i}$ being average of positions $i,j$. 

We now present the case of domain walls in two dimensions. In this scenario,
the hopping modulation associated to the domain wall
takes the form
\begin{equation}
    t_{ij} = 1 + 0.2\sin(\tilde{k}_1\mathbf{u}_{ij}\cdot \mathbf{R}_{ij}) + 0.2\sin(k_2\mathbf{u}_{ij}\cdot \mathbf{R}_{ij}),
    \label{eq:domain}
\end{equation}
where
\begin{equation}
    \tilde{k}_{1} = k_1(1+\delta\tanh(\frac{\mathbf{u}_{ij}\cdot \mathbf{R}_{ij}}{W})),
\end{equation}
The modulation above corresponds to two domain walls in the system in both $x$ and $y$ directions. Here $\delta = 0.2$ and $W = 819.2$. In Fig.~\ref{fig:square} we show the modulations on all sites of the systems $t(r_i) = \sum_{j}t_{ij}$ and compare it with the spatial distribution of magnetization obtained from mean-field calculations. In both uniform and domain wall cases, the magnetic textures closely follow the hopping modulation, demonstrating strong spatial correlation at both large and small length scales.

\section{Profiling and accuracy benchmarking}

\begin{figure*}[t!]
    \centering
    \includegraphics[width=1\linewidth]{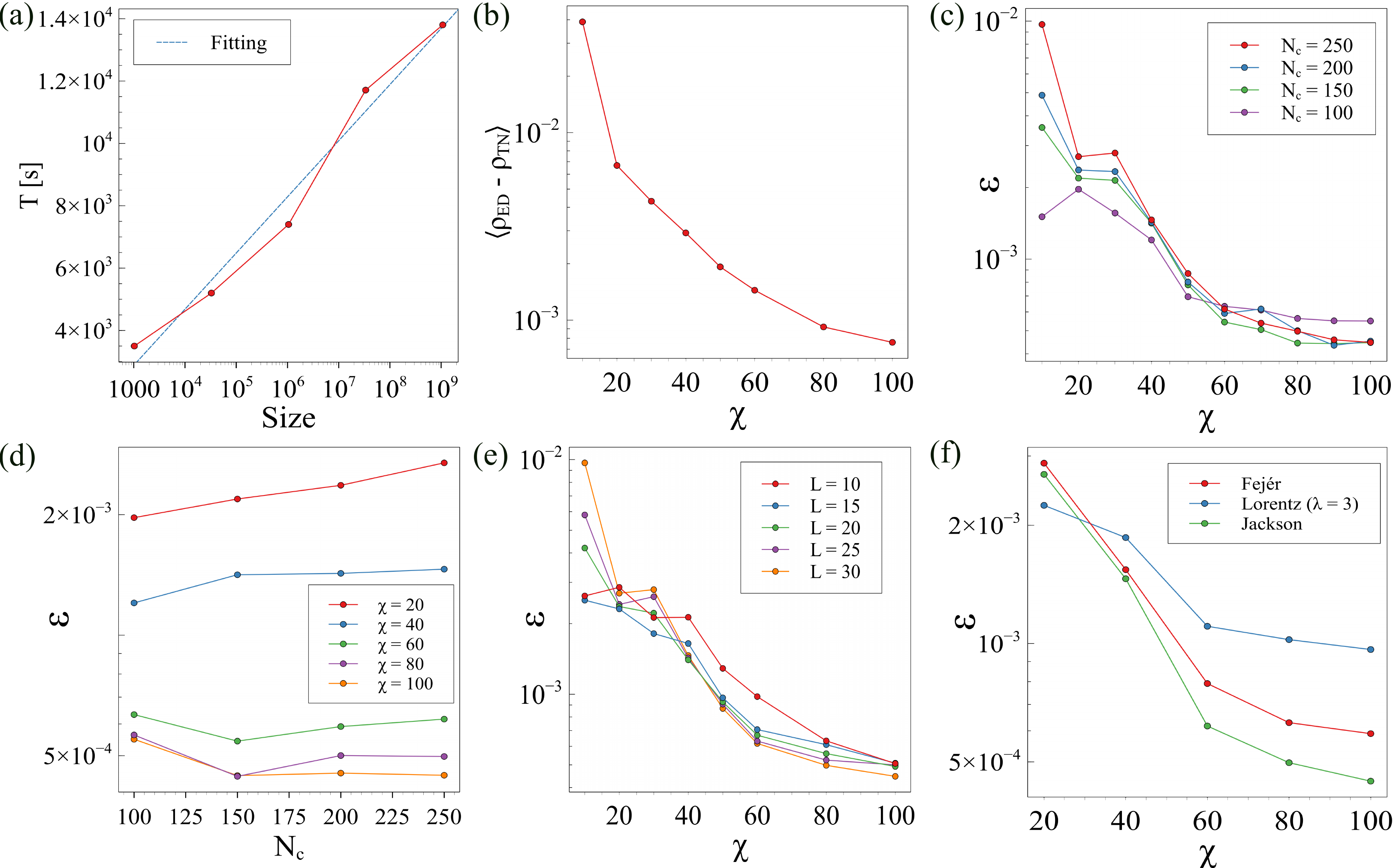}
    \caption{\textbf{Profiling and performance benchmarking of the methodology}. Panel (a) shows the total computation time for different system sizes along with a linear fit. The calculations are performed on Aalto University’s triton cluster node gpu[49-51] with the GPU Nvidia h200 and the CPU Intel Xeon Platinum 8562Y+. All KPM computations are GPU-accelerated using CUDA~\cite{Besard20192,Besard2019,itensor}. Panel (b) presents the average deviation between the electron densities obtained from the tensor-network approach and exact diagonalization (ED) for a system size of $2^{10}$. The left panels show the convergence error $\epsilon$ after 10 rounds of SCF iterations as a function of various parameters: Panel (c) shows the error versus bond dimension $\chi$ for several values of Chebyshev polynomial order $N_c$; Panel (d) shows the error versus $N_c$ for several values of $\chi$; Panel (e) shows the error versus $\chi$ for different system sizes $L$; Panel (f) shows the error versus $\chi$ for different KPM kernels. 
    Calculations are performed on
    a one-dimensional super-moir\'e system, and comparisons with ED are done for a system size of $2^{10}$ sites,
    and we took $N_c = 250$ for panel (e) and (f).
    A total of $\mathcal{N} = 10$ iterations of
    the self-consistent
    loop are performed for solving
    the self-consistent tensor network
    equations for panels (a-f).}
    \label{fig:bench}
\end{figure*}

Finally, we profile and benchmark the performance of our methodology using one-dimensional super-moir\'e systems as in Fig.~\ref{fig:bench}. 
To demonstrate the increase of
efficiency enabled by our methodology, 
we have computed the time required for solving a self-consistent problem
for different system sizes
as shown in Fig.~\ref{fig:bench}(a). We vary the system size $L$ and keep the same number of Chebyshev polynomials $N_c = 250$ and bond dimension $\chi = 100$ in all cases. The systems are governed by Hamiltonians of the same form as the systems in Fig. \ref{fig:hop} of the main text, with fixed parameters $k_1 = \frac{2\pi}{5 \sqrt{2}}$ and $k_2 = \frac{8\pi}{2^{L-1} \sqrt{3}}$.
Calculations are converged to a self-consistency error $\epsilon<10^{-3}$,
where the error in the $n$-th iteration of the mean-field procedure is
defined as $\epsilon_{n} = \text{max}(|\rho_{n} - \rho_{n-1}|)$. 

As shown in Fig.~\ref{fig:bench}(a), the total runtime $T$ for $\mathcal{N}=10$ rounds of SCF iterations exhibits a linear scaling with an exponential growth in the system size, corresponding to a logarithmic scaling with respect to the total number of lattice sites $N = 2^L$ and the size of the single-particle Hilbert space. 
For this specific problem and at fixed $\chi$ and $N_c$,
our approach achieves $\mathcal{O}(\log N)$ computational complexity, a dramatic improvement over traditional KPM approaches including our previous sparse-matrix implementation which typically scale as $\mathcal{O}(N^2)$~\cite{RevModPhys.78.275,Fumega2024}.
We note that this scaling appears
from keeping the bond dimension $\chi$ and number of Chebyshev
polynomials $N_c$ fixed, which in particular
enables having a self-consistency error $\epsilon<10^{-3}$ for all system sizes. 
In case the parameters $\chi$ and $N_c$ were dynamically adjusted to target a specific
accuracy, the scaling with the system size may be different.
We note that the number of Chebyshev polynomials
enforces an effective energy smearing~\cite{RevModPhys.78.275}, and dynamically updating that may introduce instabilities in the self-consistent loop.
The choice of fixing $\chi$ and $N_c$ in the self-consistent loop is similar to fixing the maximum kinetic energy of plane waves
and the Fermi temperature in density functional theory calculations~\cite{Giannozzi2017},
parameters that are most commonly fixed during the self-consistent loop and are
not dynamically updated.
It is worth noting that once a solution is converged with a certain $\chi$ and $N_c$, the self-consistent loop could be restarted with the 
converged solution using a higher value for $\chi$ and $N_c$.
The approximate $\mathcal{O}(\log N)$ complexity
directly leads to speed-up of several orders of
magnitude for systems sizes above a million, making it possible to perform self-consistent calculations on systems with up to billions of sites, which were previously intractable. 
It is finally worth noting that
the motivation for using tensor networks is not solely the reduction of computational time. Rather, the primary advantage lies in enabling calculations for extremely large systems where storing even a single sparse Hamiltonian matrix is infeasible due to memory constraints. By compressing the Hamiltonian into an MPO, we drastically reduce the memory cost and thus make it possible to store and manipulate the Hamiltonians at this scale.

For the accuracy benchmarking, as our algorithm
fully relies on tensor networks, the bond dimension directly controls the fidelity of the MPO approximation. We took a bond dimension $\chi=100$ for all calculations for one- and two-dimensional systems shown in the main text and SM. For one-dimensional systems, self-consistent errors were typically below $10^{-3}$, while for the more challenging two-dimensional cases, errors remained within $2\times10^{-3}$, which are typical values of self-consistency error required to reach convergence in mean-field calculations in
tight binding models~\cite{PhysRevLett.99.177204,Yazyev2010}.
To further prove that
a bond dimension of $\chi  = 100$ ensures good precision in our calculations,
we have benchmarked the performance of our method using one-dimensional super-moir\'e systems as in Fig.~\ref{fig:bench}. Here we analyze how the key parameters such as the bond dimension $\chi$, the number of Chebyshev moments $N_c$, and the choice of KPM kernel influence the performance of the SCF calculations based on convergence error $\epsilon$ after $\mathcal{N}=10$ rounds of iterations.  
First, we benchmark the accuracy of our method by directly comparing the results with those obtained via exact diagonalization (ED) for a large system. $N_c = 250$ is used for our algorithm. As shown in Fig.~\ref{fig:bench}(b), the average deviation of electron densities between our tensor network approach and ED systematically decreases with increasing bond dimension $\chi$ as expected. Notably, the error drops below $10^{-3}$ for moderate bond dimensions, demonstrating that our method achieves very good quantitative agreement with ED.
This result confirms that our tensor network compression introduces only minimal approximation error when the bond dimension is chosen appropriately, validating the reliability of our approach for large-scale simulations.
Furthermore, a plateauing behavior of the error curve appears with for increasing $\chi$, especially when $\chi > 60$. Further increasing $\chi$ will not improve the precision, 
as in this region the $N_c$ is limiting the precision. 
This stems from the energy scale introduced by
$N_c$, as
$\delta_c \sim 1/N_c$ in the problem~\cite{RevModPhys.78.275}, similar to an effective smearing in the spectral functions, whose impact
will be observed once the self-consistency error becomes comparable or small than $\delta_c$. Therefore,
in order for KPM to agree with ED at those errors, a larger $N_c$ will be needed. 

For Figs.~\ref{fig:bench}(c-f), the left panels show the convergence error $\epsilon$ after $\mathcal{N}=10$ SCF iterations as a function of various parameters, where for the $n$th iteration $\epsilon_n= |\rho_n - \rho_{n-1}|$, $\rho_n$ being
the electron densities at $n$th iteration.
Figs.~\ref{fig:bench}(c,d) illustrate how $\chi$ and $N_c$ affect the convergence behavior for a system with $2^{30}$ sites. Interestingly, we observe that when $\chi$ is small, increasing $N_c$ leads to worse convergence (Figs.~\ref{fig:bench}(d)), 
which can be rationalized as follows. 
If the number of polynomials $N_c$ is increased to arbitrarily large values,
the required bond dimension is likely to increase, in particular because the
largest Chebyshev polynomial may require a large bond dimension to be
properly captured.
Specifically, an insufficient bond dimension gives rise to a reconstructed
density of states that becomes negative at certain energies if $N_c$ is too large. 
Therefore, the
positivity of the density of states can be used as a proxy signaling that the bond dimension
$\chi$ is large enough for a certain number of polynomials $N_c$.
For sufficiently large $\chi$, the convergence error can
be made below a desired threshold
for a wide range of number of polynomials $N_c$.
These results highlight the importance of a balanced choice between $N_c$ and $\chi$: a high $N_c$ is only beneficial when the bond dimension is large enough for the MPOs to correctly approximate the polynomial components, while larger $\chi$ will lead to heavy increase in computational time consumption. 
It finally worth noting that $N_c$ controls the energy smearing in the calculation~\cite{PhysRevB.87.195448}, analogous to temperature in experiments. As a result, very large $N_c$ values are only necessary when resolving correlated states with very small energy scales. The interplay between $N_c$ and $\chi$ will thus deserves
a focused detailed study on its own, 
and we hope our findings will motivate further exploration in this direction.

Fig.~\ref{fig:bench}(e) presents the convergence error as a function of the bond dimension $\chi$ for systems of varying sizes
and after $\mathcal{N} = 10$ iterations.
The minimum error $\epsilon$
is thus bounded by the number of iterations $\mathcal{N}$.
We solve our self-consistent tensor network problem
with a linear mixing scheme,
and therefore the self-consistency error reached after $\mathcal{N}$
iterations scales,
in a best case scenario,
approximately as $\epsilon_{\mathcal{N}}=e^{-\lambda \mathcal{N}}$~\cite{PhysRevB.54.11169},
where $\lambda$ depends on the mixing between new and old tensor networks.
The number of iterations ${\mathcal{N}}$ was fixed at ${\mathcal{N}}=10$
because this
choice enables to reach errors below $\epsilon \equiv \epsilon_{{\mathcal{N}}=10} \sim 10^{-3}$
in our reference calculations.
As a result, Fig.~\ref{fig:bench}(e) 
highlights whether all different system sizes are able to reach
convergence
at a similar rate after $\mathcal{N} = 10$ iterations, and therefore enables quantifying
if the
finite bond dimension considered allows reaching a self-consistent solution after
those iterations. As it is observed, all system
sizes show a comparable convergence. It is also worth noting that for $L=10$, bond dimension
$\chi=100$ corresponds to a fully covered Hilbert space, and therefore
that calculation is equivalent to a full matrix implementation of the
KPM self-consistent algorithm.
As expected, at small $\chi$, smaller systems exhibit lower errors than larger ones, consistent with their more compact representation. The key observation occurs at $\chi = 100$. Beyond this value $\chi = 100$, the convergence error becomes nearly system-size independent, with exceptionally large systems achieving an accuracy comparable to small ones. This signifies that the tensor network has entered a regime where it faithfully represents the self-consistent solution. We note that more complex Hamiltonians may require a bigger bond dimension to reach the convergence level achieved in our calculations. An error $\epsilon$ below $10^{-3}$ at $\chi = 100$ is the typical
value required for self-consistent studies of symmetry-broken states in graphene~\cite{PhysRevLett.99.177204,Yazyev2010}, and therefore it is fitting for our purposes. The fact that we achieve this precision self-consistently across system sizes up to a billion sites demonstrates that our chosen bond dimension is sufficient to reach
a satisfactory convergence in our regime. 

Finally, we compare the convergence performance of different kernel functions used in the KPM expansion~\cite{RevModPhys.78.275}. Throughout this work, we primarily adopt the Jackson kernel~\cite{Jackson1912}, which effectively suppresses Gibbs oscillations arising from the Chebyshev truncation. It is defined as:
\begin{equation}
    g_{n}^{J} = \frac{(N_c-n+1)\cos(\frac{n\pi}{N_c+1}) + \sin(\frac{n\pi}{N_c+1})\cot(\frac{\pi}{N_c+1}) }{N_c+1}.
\end{equation}
By applying $g_{n}$ to $n$th Chebyshev polynomials $T_n(\mathcal{H})$, the Gibbs oscillations from truncation of KPM are suppressed. Although Jackson kernel is the most popular choice in KPM calculations, to assess whether it still remains the optimal choice in the context of our methodology, we benchmark it against two widely used alternatives: The Fej\'er kernel 
\begin{equation}
    g_{n}^{F} = 1- \frac{n}{N_c},
\end{equation}
and the Lorentz kernel
\begin{equation}
    g_{n}^{L} = \frac{\sinh[\lambda(1-\frac{n}{N_c})]}{\sinh(\lambda)},
\end{equation}
here $\lambda$ is a parameter used for controlling the smoothness and we use $\lambda = 3$. As shown in  Fig.~\ref{fig:bench}(f), the Jackson kernel clearly outperforms the other two, providing the lowest convergence error across all settings. The Fej\'er kernel performs moderately well, while the Lorentz kernel exhibits the worst convergence, likely due to its limited ability to suppress high-order oscillations. This comparison confirms that Jackson kernel remains the optimal choice for our approach for performing KPM calculations for tensor network compressed super-moir\'e Hamiltonians. 
 
%\bibliography{biblio}

%

\end{document}